\documentclass[pre,aps,twocolumn,amsmath,amssymb]{revtex4-1}

\pdfoutput=1
\usepackage{hyperref}
\usepackage{graphicx}

\def\cal#1{\mathcal{#1}}
\def\eqq#1{Eq.~(\ref{#1})}
\def\eqs#1{Eqns.~(\ref{#1})}
\def\eq#1{(\ref{#1})}

\def\f#1{Fig.~\ref{#1}}

\def\fs#1{Figs.~\ref{#1}}

\def\s#1{Section~\ref{#1}}

\def\c#1{~\cite{#1}}
\def\csupp#1{~\cite{#1}}

\def\beq{\begin{equation}}
\def\eeq{\end{equation}}
\def\bea{\begin{eqnarray}}
\def\eea{\end{eqnarray}}

\def\kt{k_{\rm B}T}

\begin{document}

\title{Emergent rhombus tilings from molecular interactions with $M$-fold rotational symmetry}

\author{Stephen Whitelam$^{1}$}\email{{\tt swhitelam@lbl.gov}}
\author{Isaac Tamblyn$^{2}$}
\author{Juan P. Garrahan$^{3}$}
\author{ Peter H. Beton$^{3}$}

\address{$^1$Molecular Foundry, Lawrence Berkeley National Laboratory, 1 Cyclotron Road, Berkeley, CA 94720, USA\\
$^4$Department of Physics, University of Ontario Institute of Technology, Oshawa, Ontario L1H 7K4, Canada\\
$^3$School of Physics and Astronomy, University of Nottingham, Nottingham NG7 2RD, UK}

\begin{abstract}

We show that model molecules with particular rotational symmetries can self-assemble into network structures equivalent to rhombus tilings. This assembly happens in an emergent way, in the sense that molecules spontaneously select irregular 4-fold local coordination from a larger set of possible local binding geometries. The existence of such networks can be rationalized by simple geometrical arguments, but the same arguments do not guarantee networks' spontaneous self-assembly. This class of structures must in certain regimes of parameter space be able to reconfigure into networks equivalent to triangular tilings.
\end{abstract}

\maketitle

The self-assembly of molecules at surfaces has practical application, because it allows us to influence the properties of that surface\c{elemans2009molecular,bartels2010tailoring}. It is also of fundamental interest: self-assembly at surfaces has provided insight into the general question of how to encourage the self-assembly of a desired structure while avoiding the self-assembly of undesired structures\c{whitesides2002self,philp1996self,hagan2006dynamic,wilber2007reversible,rapaport2010modeling}. In several cases, basic considerations of molecular symmetry and geometry have been shown to be key to understanding the outcome of self-assembly at surfaces. For example, a wide range of molecules whose atomic details differ but whose interactions possess three-fold rotational symmetry form honeycomb networks and polygon variants thereof, both in and out of equilibrium\c{geim2007rise,he2005self,lichtenstein2012crystalline,bieri2009porous,palma2010atomistic,whitelam2014common}. Molecules whose binding geometries are equivalent to those of rhombus tiles form rhombus tilings of the plane\c{blunt2008random,blunt2008directing,stannard2010entropically,whitelam2012random}.

In those examples, the restriction of preferred molecular binding geometries to {\em only} those characteristic of a particular network encourages formation of that network, and helps prevent the formation of other possible networks. Here we demonstrate in simulations an example of network self-assembly that happens instead in an {\em emergent} fashion. Collections of mutually attractive model molecules having particular rotational symmetries spontaneously select only a sub-set of possible local binding geometries. By doing so, they self-assemble into extended networks equivalent to rhombus tilings of the plane\c{fisher1961statistical,kasteleyn1963dimer,blšte1982roughening,henley1991quasicrystals,destainville1998entropy,destainville2005random,alet2006classical}.

\begin{figure}[h!] 
   \centering
 \includegraphics[width=\linewidth]{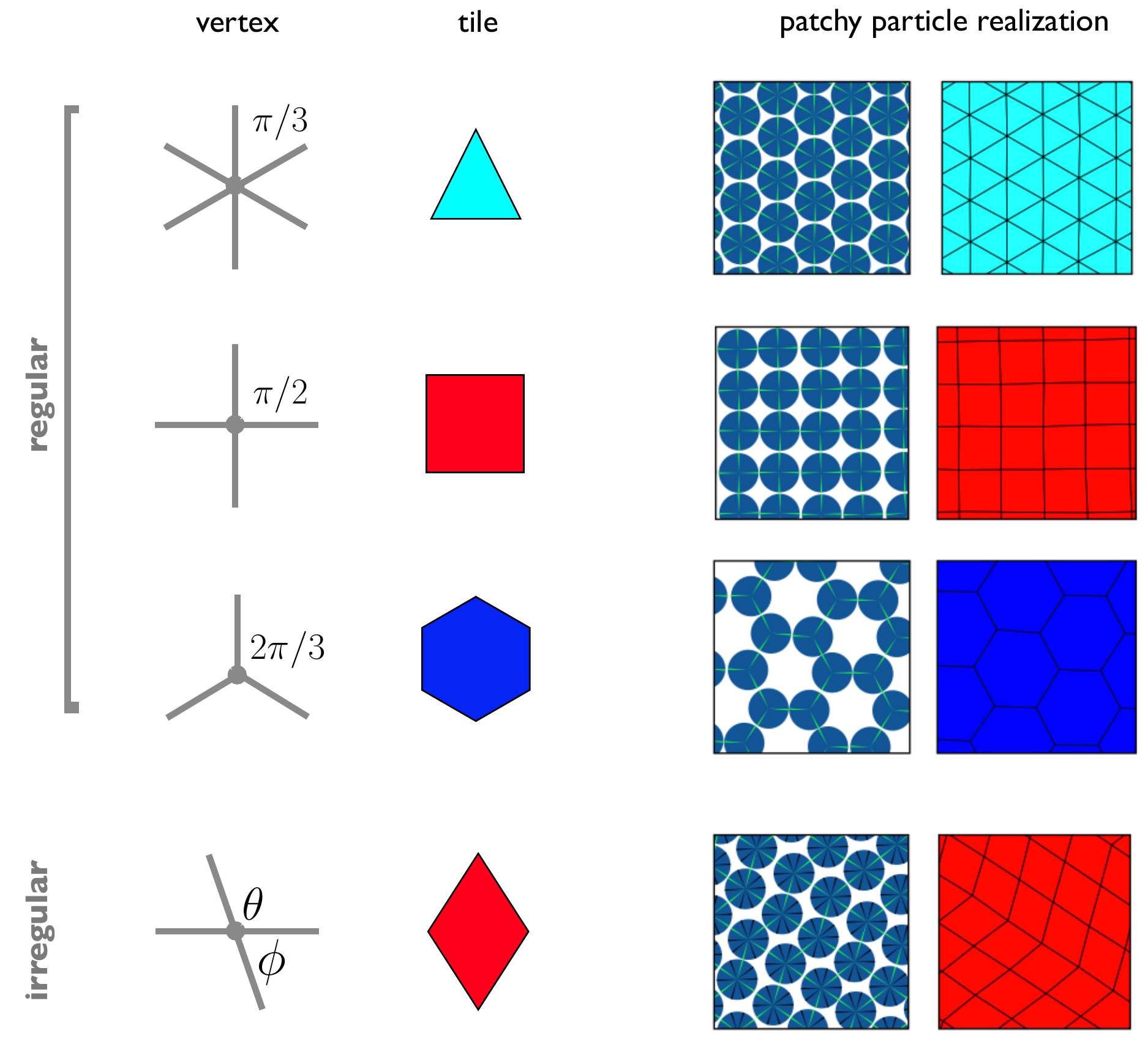} 
   \caption{\label{fig1} Platonic tilings of triangles, squares or hexagons result from regular 6-vertices, 4-vertices or 3-vertices, respectively, and can self-assemble from collections of building blocks with the corresponding rotational symmetries\c{doppelbauer2010self,antlanger2011stability,doye2007controlling}. The 4-vertex can be made irregular and result in a tile, a rhombus, whose sides are of equal length. As we show in this paper, irregular 4-vertex networks equivalent to rhombus tilings spontaneously self-assemble from building blocks with $M$-fold rotational symmetry, where $M \geq 4$ is even and not a multiple of 6.}
\end{figure}
\begin{figure*}[ht!] 
   \centering
 \includegraphics[width=0.8\linewidth]{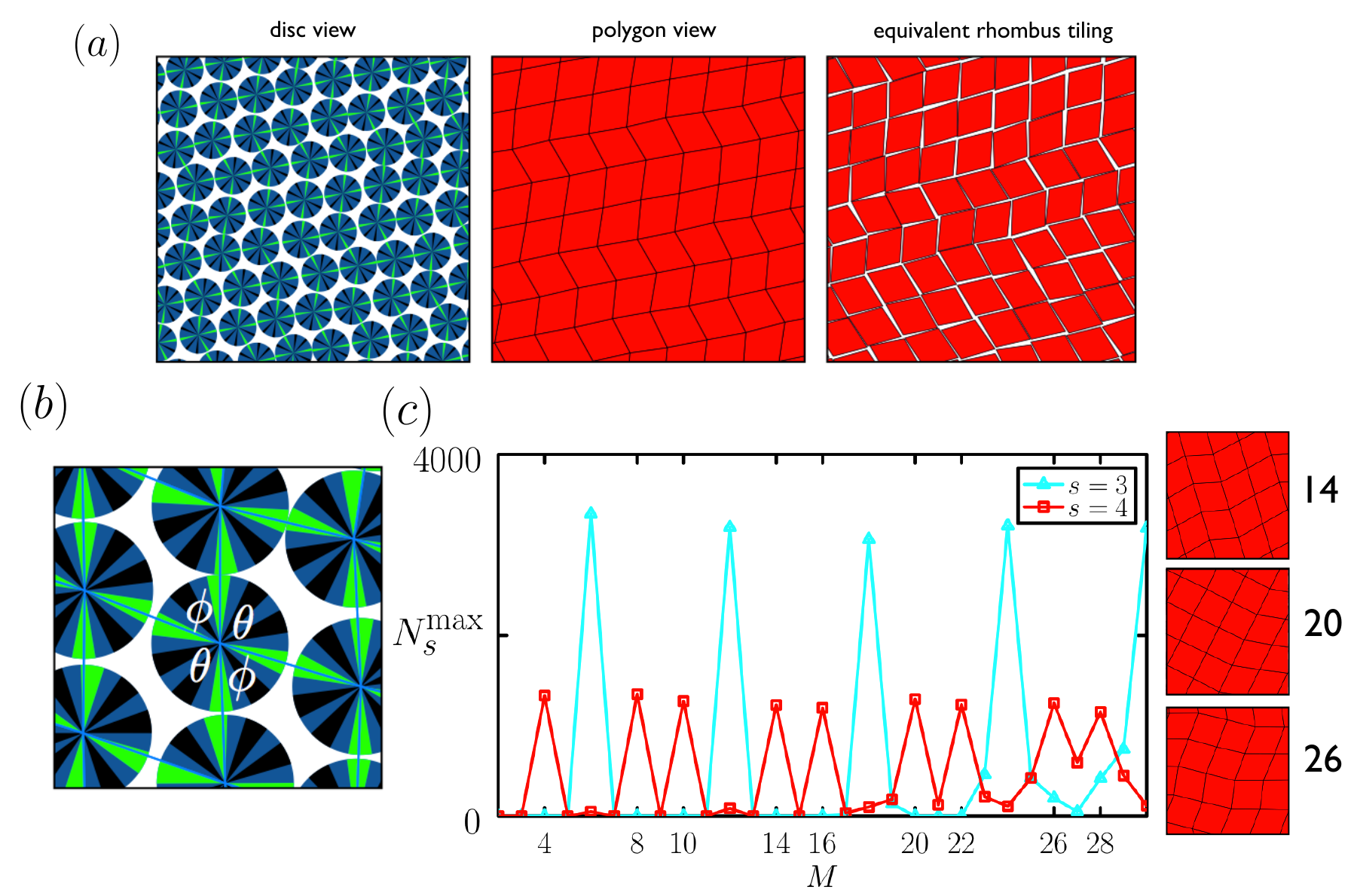} 
   \caption{\label{fig2} Rhombus tilings emerge from particular regular rotational symmetries of building block interactions. (a) Region of rhombus tiling self-assembled from $10$-patch discs ($w=5^\circ$), in disc (left) and polygon (center) views. This tiling is similar to the `tilted' rhombus phase\c{blunt2008random,stannard2010entropically,ye2013competition} (right) that results from the self-assembly of mutually attractive rhombus tiles with internal angles $2\pi/5$ and $3 \pi/5$. (b) Geometry used to motivate the observation that unstrained rhombus tilings can form from regular $M$-patch discs, where $M$ is even and not a multiple of 6. (c) In simulations, discs of this nature spontaneously self-assemble into rhombus tilings if cooled slowly. Here we plot the largest number of convex polygons of size $s=3$ or 4 obtained upon cooling collections of $M$-patch discs with $w=2^\circ$. Inspection of configurations reveals the 4-gons at the labeled values of $M$ to be rhombuses (see snapshots right and \fs{fig_array} and \ref{fig_array2}).}
\end{figure*}

To provide context for this result we consider in \f{fig1} the regular Platonic tilings, which are coverings of the plane by non-overlapping regular triangular, square, and hexagonal tiles\c{grfinbaum1987tilings}. The vertices of such tilings possess 6-fold, 4-fold, and 3-fold rotational symmetry, respectively. Other authors have shown that the networks equivalent to such tilings can be realized by the self-assembly of model molecules whose interactions possess the appropriate rotational symmetries\c{doppelbauer2010self,antlanger2011stability,doye2007controlling}. By way of illustration, we show pictures in the figure of such networks, and the equivalent tilings, self-assembled in our simulations from collections of `patchy particles'\c{zhang2004self,tartaglia2010association} in two dimensions. The particles in question are hard discs that possess $M$-fold rotational symmetry of their interactions: discs attract each other in a pairwise fashion via $M$ regularly-placed `patches' of opening angle $2w$. Two engaged patches result in a favorable binding energy $-\epsilon \, \kt$. Disc interactions are `square-well' in both angle\c{kern2003fluid} and range; see \s{sec_model}. By slowly cooling collections of such discs (\s{sec_cooling}) we observe the spontaneous assembly of the networks shown.

Thermodynamically, it is clear why discs form these networks and not other ones. Discs within these networks have all their patches engaged, and so possess the least possible energy. Networks are also geometrically unstrained -- on average, disc patch bisectors point along the lines connecting vertices -- and so discs retain as much rotational and vibrational entropy as possible. Lack of strain is possible because the rotational symmetries of the network vertices and disc interactions are identical, and because the resulting tile has all sides of equal length, enabling its formation by discs possessing a single preferred (average) interaction distance.  Dynamically, assembly is successful because slow cooling allows discs to select low-energy local binding environments before interactions become strong enough to cause kinetic trapping.

Given that the networks equivalent to the Platonic tilings are built from regular vertices and tiles, it is natural to ask if one can arrange for the self-assembly of irregular variants of these networks. The 6-vertex, if deformed, gives rise (locally) to triangles whose sides are not all of equal length. A single type of molecule could therefore not form a network of such vertices without strain, unless it possessed multiple preferred interaction ranges (see e.g.\c{rechtsman2006designed,rechtsman2005optimized}). But the 4-vertex is different. As shown in \f{fig1}, a 4-vertex harboring two types of internal angle, $\phi$ and $\psi$, is equivalent to a rhombus tile possessing these two internal angles. A rhombus has all sides equal, and so, in principle, a single model molecule with the appropriate rotational symmetry and a single preferred interaction range could form an unstrained network equivalent to a tiling of such rhombuses. 

This is what we show in the lower right-hand panels of \f{fig1}. However, the model molecule in question possesses not irregular 4-fold rotational symmetry but regular 10-fold rotational symmetry. Cooling collections of this molecule (with patches of half-width $w=5^\circ$) results in self-assembly of the pattern shown. In this pattern, discs engage only 4 of their 10 patches: the irregular 4-vertex has emerged spontaneously. By comparing the left and center panels of \f{fig2}(a), which show a larger piece of this network, we see that domains of parallel-pointing rhombuses result when discs engage 4 patches whose bisectors are separated by angles $(2,3,2,3) \times \pi/5$, in order. Boundaries between parallel-pointing domains result when engaged patch bisectors possess angular separations ordered instead as $(2,2,3,3) \times \pi/5$. The rhombus tiling that results is similar (see right panel) to that made by the self-assembly of rhombus-shaped tiles possessing internal angles $2\pi/5$ and $3 \pi/5$ (\s{sec_rhombus}).

Thus, collections of model molecules with 10-fold rotational symmetry spontaneously select local irregular 4-fold coordination and so self-assemble into an extended network. This example turns out to be a particular case of a more general phenomenon, as we now describe. We can ask from what regular rotational symmetries are rhombus tilings possible geometrically, focusing on the narrow-patch limit in which patch bisectors must point along the lines connecting network vertices. In \f{fig2}(b) we show a picture of a 10-patch disc that forms the vertex of a self-assembled rhombus tiling (for the purposes of the following argument we shall ignore that fact that disc patches in this picture are of finite width). The angles $\phi$ and $\theta$ are the two internal angles of the rhombus. Because these angles meet at a vertex, they satisfy
\beq
\label{constraint}
2(\phi + \theta) = 2 \pi.
\eeq
Imagine now that this disc has not 10 but $M$ patches, and let the angles $\phi$ and $\theta$ correspond to the intervals separating an integer numbers of patches, say $\alpha$ and $\beta$, respectively. That is, let $\phi = 2 \pi \alpha /M$ and $\theta = 2 \pi \beta /M$. Inserting these expressions into \eqq{constraint} gives 
\beq
\label{condition}
\alpha+\beta=M/2.
\eeq
Because $\alpha$ and $\beta$ are integers, $M/2$ must be an integer, and so $M$ must be even. Thus, it is geometrically possible for discs with even-numbered rotational symmetry to form unstrained rhombus tilings. Thinking now of the potential self-assembly of such tilings, we note that if $M$ is a multiple of 6 then discs can also form an unstrained triangular tiling, and this -- being made from 6-vertices rather than 4-vertices -- will be energetically preferred to the rhombus tiling. Simple geometric arguments therefore suggest the possibility of having rhombus tilings self-assemble from model molecules with $M$-fold rotational symmetry, where $M (\geq 4)$ is even and not a multiple of 6.

In \f{fig2}(c) we show using simulations of $M$-patch discs that this possibility can be realized: collections of such molecules {\em do} self-assemble into rhombus tilings if cooled slowly. We plot the largest number $N_s^{\rm max}$ of polygons (drawn atop networks) having $s$ sides that were seen over the course of long cooling simulations. Networks self-assembled at the labeled values of $M$ contain numerous 4-gons. Inspection of networks (see e.g. snapshots right and \fs{fig_array} and \ref{fig_array2}) reveals these 4-gons to be rhombuses, arranged into rhombus tilings. The tilings predicted by geometry to exist are therefore kinetically accessible: discs self-assemble into rhombus tilings in preference to any of the many possible disordered networks.

Tilings are imperfect, containing grain boundaries that result from the dynamic process of self-assembly (the more so as $M$ increases). Rhombuses in tilings can also be of more than one shape. This observation is consistent with the fact that \eqq{condition} constrains only the sum of $\alpha$ and $\beta$: an $M$-patch disc can make rhombuses of more than one type $(\alpha,M/2-\alpha)$, by which we mean a rhombus with internal angles $(\alpha,M/2-\alpha) \times 2\pi/M$. Multiple rhombus types are indeed seen in simulations for $M$ sufficiently large. For instance, the 10-patch disc makes only the (2,3) rhombus; the (1,4) rhombus is forbidden sterically. But for 16 patches and upwards, rhombuses in tilings are of more than one type. For example, the 20-patch disc forms the $(4,6)$ and $(5,5)$ rhombuses (a boundary between these tile types is shown in \f{fig2}(c)'s middle snapshot); the 28-patch disc forms the $(7,7)$, $(9,5)$ and $(8,6)$ rhombuses.
\begin{figure*}[ht] 
   \centering
 \includegraphics[width=0.8\linewidth]{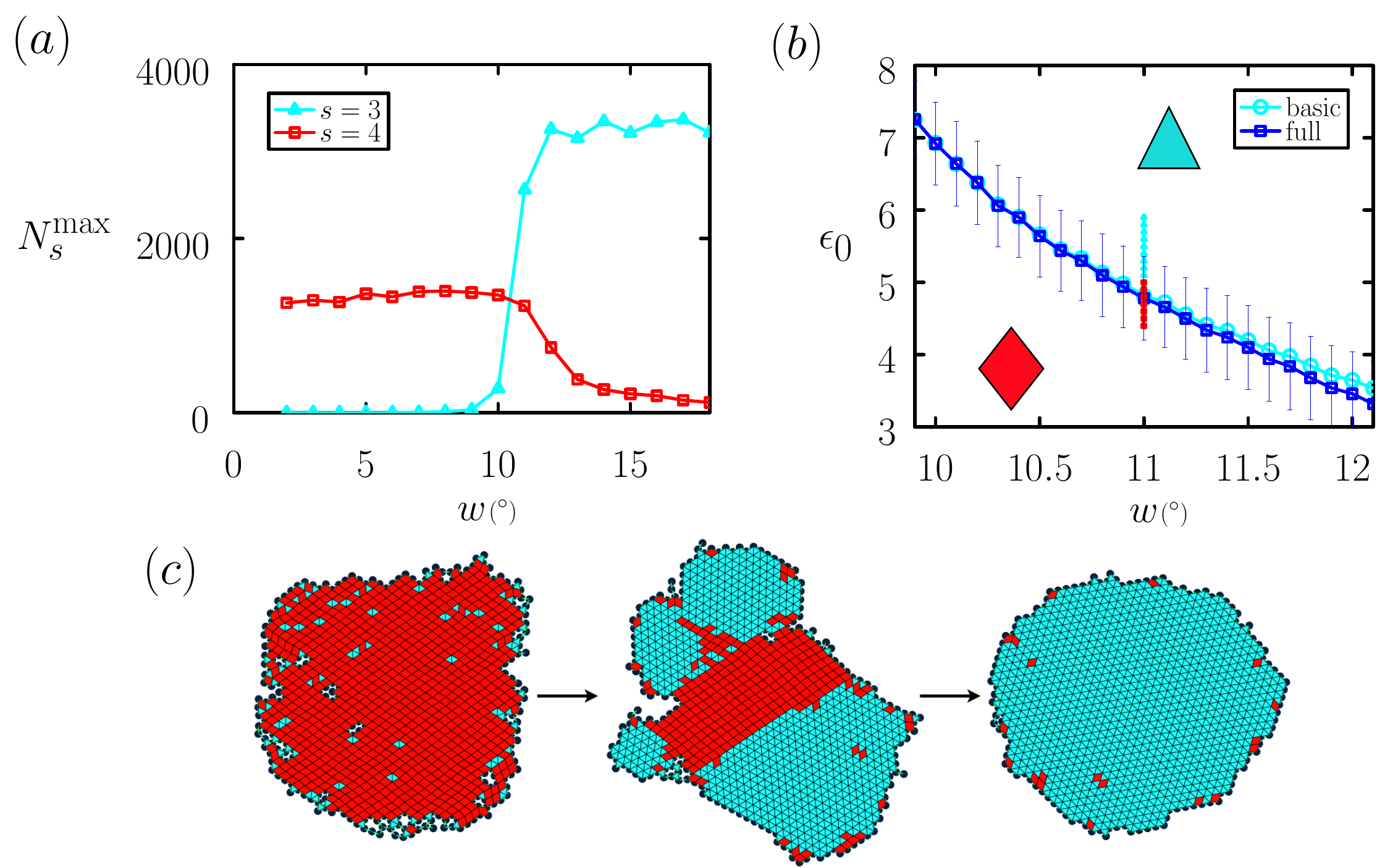} 
   \caption{\label{fig3} Rhombus-triangle polymorphism. (a) Cooling collections of 10-patch discs results in self-assembly of the rhombus phase when interactions are orientationally specific, and of the triangle phase as interactions approach the isotropic limit. Between these limits the two phases can coexist and interconvert, with rhombuses favored by entropy and triangles by energy (see also \f{fig_loop_counts}). Panel (b) shows the rhombus-triangle phase boundary calculated at zero pressure using mean-field theory. The lines labeled `full' and `basic' are the predictions of \eqq{complex} and \eqq{simple}, respectively. Small symbols for $w=11^\circ$ describe the results of direct coexistence simulations, which are consistent with the mean-field estimate (see \s{sec_stability}). Panel (c) shows a rhombus-to-triangle transformation at parameter set $(\epsilon=7, w=11^\circ$), where \eqq{complex} predicts the bulk free-energy density of the triangle phase to be $2.2 \, \kt$ less than that of the rhombus one.}
\end{figure*}

These emergent rhombus tilings must be polymorphic with at least other solid phase in certain regimes of parameter space. To see this, we note than in simulations thus far we have focused on model molecules with specific orientational interactions, i.e. narrow `patches', because rhombus tilings are free of geometrical strain and so can in principle form from discs whose patches are of infinitesimal width (we checked that the expected rhombus tilings self-assemble, for $M \leq 32$, for the finite but narrow half-width $w=0.5^\circ$). However, real molecules possess finite flexibility of binding, and so are better modeled by discs whose patches have appreciable width\c{saika2013understanding,whitelam2014common}. In the limit of wide patches, when the whole circumference of the disc is sticky (i.e. when $w=\pi/M$), networks equivalent to triangular tilings must result. Somewhere between the wide- and narrow-patch limits, then, polymorphism must occur. 

We verified that this is so for 10-patch discs. In \f{fig3}(a) we show that cooling collections of 10-patch discs with patches of half-width $w \lesssim 10^\circ$ results in tilings of 4-gons (which we verified by eye to be rhombuses), while discs with wider patches form triangular tilings. $N_s^{\rm max}$ again denotes the largest number of $s$-gons observed over the course of long cooling simulations. Triangular tilings for $w\approx 10^\circ$ are irregular, and become regular as $w$ approaches $\pi/10=18^\circ$. Near $w = 10^\circ$ we observe polymorphism: the first tiling to assemble is the rhombus one, which, as temperature is reduced (as $\epsilon$ is increased), converts to a triangular tiling. 

This conversion occurs because discs in networks equivalent to rhombus tilings are unstrained geometrically, and so possess more rotational and translational entropy than do discs in strained triangular tilings. But discs forming triangular tilings make 6 contacts rather than 4, and so are energetically preferred. Triangles therefore emerge at low temperature. A related energy-versus-entropy competition between open and close-packed lattices has been demonstrated for patchy silica spheres\c{mao2013entropy}. In \f{fig3} we show the phase boundary between triangle and rhombus tilings, which we calculated at zero pressure using a mean-field theory whose parameters were informed by simulation (see \s{sec_stability}). Above the line $\epsilon_0(w)$, triangular tilings are lower in free energy than rhombus tilings~\footnote{The value of $w$ at which polymorphism occurs decreases with increasing $M$. For 10-patch discs we see polymorphism for $w\approx 10^\circ$; for 32-patch discs this polymorphism occurs at smaller $w$: discs self-assemble into rhombus tilings when  $w=0.5^\circ$, but self-assemble into triangular tilings when $w=2^\circ$ (the value used to produce \f{fig2}(c)).}. Panel (c) shows a rhombus-to-triangle tiling transformation involving discs of patch half-width $w=11^\circ$. We therefore predict that emergent rhombus tilings formed by molecules with sufficient flexibility of binding should reconfigure in response to changes of temperature.

We note finally that the Archimedean tilings, which are made from regular vertices and harbor two tile types, cannot be prepared in an unstrained manner using discs with regular $M$-fold symmetry (i.e. cannot be prepared in the narrow patch limit). All 5-vertex and 4-vertex Archimedean tilings contain equilateral triangles. To make equilateral triangles from an $M$-patch disc in an unstrained way requires $M$ to be a multiple of 6, in which case discs self-assemble into the energetically preferred 6-vertex triangular tiling; see \f{fig2}(c). Self-assembly of Archimedean tilings can be achieved instead using irregular patch placement\c{antlanger2011stability}, or an odd number of suitably wide patches\c{doye2007controlling,van2012formation,PhysRevLett.110.255503} (or the equivalent effective coordination\c{ecija2013five,millan2014self}); see \f{fig5}.

We have shown that model molecules with particular rotational symmetries can self-assemble into rhombus tilings in an emergent way, by spontaneously selecting irregular 4-fold local coordination from a larger set of possible local binding motifs. The existence of rhombus tilings from $M$-fold-coordinated discs, where $M \geq 4$ is even and not a multiple of 6, can be rationalized by simple geometric arguments (see \eqq{condition}). However, the self-assembly of such tilings does not follow automatically, because it requires discs to avoid forming other possible network structures, and these are numerous when $M$ is large. Nonetheless, our simulations show rhombus tilings to spontaneously self-assemble from discs with the appropriate rotational symmetries. Given the fine control displayed in numerous examples of molecular synthesis for the purposes of self-assembly at surfaces\c{elemans2009molecular,bartels2010tailoring}, it is likely that there are ways of realizing the appropriate rotational symmetries experimentally, and therefore realizing the rhombus tilings identified here. We predict that such tilings, if made from molecules with sufficient binding flexibility, could reconfigure in response to changes of temperature, which may have energy-transfer or phase-change application.

{\em Acknowledgements.} This work was done as part of a User Project at the Molecular Foundry, Lawrence Berkeley National Laboratory, supported by the Director, Office of Science, Office of Basic Energy Sciences, of the U.S. Department of Energy under Contract No. DE-AC02--05CH11231. J.P.G. and P.H.B. acknowledge support from EPSRC Grant no. EP/K01773X/1. I.T. acknowledges support from NSERC.

%\bibliography{bib}

%

\onecolumngrid
\clearpage

~\\
\begin{center}
{\noindent\Large{Supplementary Information}}\\
\noindent \Large for\\
{\noindent \Large ``Emergent rhombus tilings from molecular interactions with $M$-fold rotational symmetry''}\\
~\newline
{\noindent \normalsize Stephen Whitelam$^{1,*}$, Isaac Tamblyn$^{2}$, Juan P. Garrahan$^{3}$, Peter H. Beton$^{3}$}\\
{ \small
~\\
\noindent$^1$ Molecular Foundry, Lawrence Berkeley National Laboratory, 1 Cyclotron Road, Berkeley, CA, USA\\
\noindent$^2$ Department of Physics, University of Ontario Institute of Technology, Oshawa, Ontario L1H 7K4, Canada\\
\noindent$^3$School of Physics and Astronomy, University of Nottingham, Nottingham NG7 2RD, UK\\
\noindent$^*${\tt swhitelam@lbl.gov}}

\end{center}

\vspace{-0.5cm}

\renewcommand{\theequation}{S\arabic{equation}}
\renewcommand{\thefigure}{S\arabic{figure}}
\renewcommand{\thesection}{S\arabic{section}}

\setcounter{equation}{0}
\setcounter{section}{0}
\setcounter{figure}{0}

\setlength{\parskip}{0.25cm}%
\setlength{\parindent}{0pt}%

\section{Disc model}
\label{sec_model}

Our disc model consists of hard `patchy' discs of diameter $a$. Discs live in continuous space on a featureless two-dimensional substrate. Discs are decorated by $M$ stripes or `patches', shown black (or green) in \f{fig_model}, each of which is a sector of opening angle $2w$. Patches are equally spaced in angle around the disc center, i.e. neighboring patch bisectors are separated by an angle $2 \pi/M$. Discs bind in a pairwise fashion, with favorable energy of interaction $-\epsilon \, \kt$, if two disc centers lie within a distance $a+\Delta$, where $\Delta=a/10$, and if those discs' center-to-center line cuts through one patch on each disc\csupp{kern2003fluid}. Patches engaged in this manner are shown green in figures. To ensure that a patch can bind to only one other patch, we require $w<\arcsin\left[(a/2)/(a+\Delta)\right]= \arcsin(5/11) \approx 27.0^{\circ}$. Note that the entire circumference of the disc is sticky when $2w = 2 \pi/M$.

\begin{figure}[ht!] 
   \centering
 \includegraphics[width=0.5\linewidth]{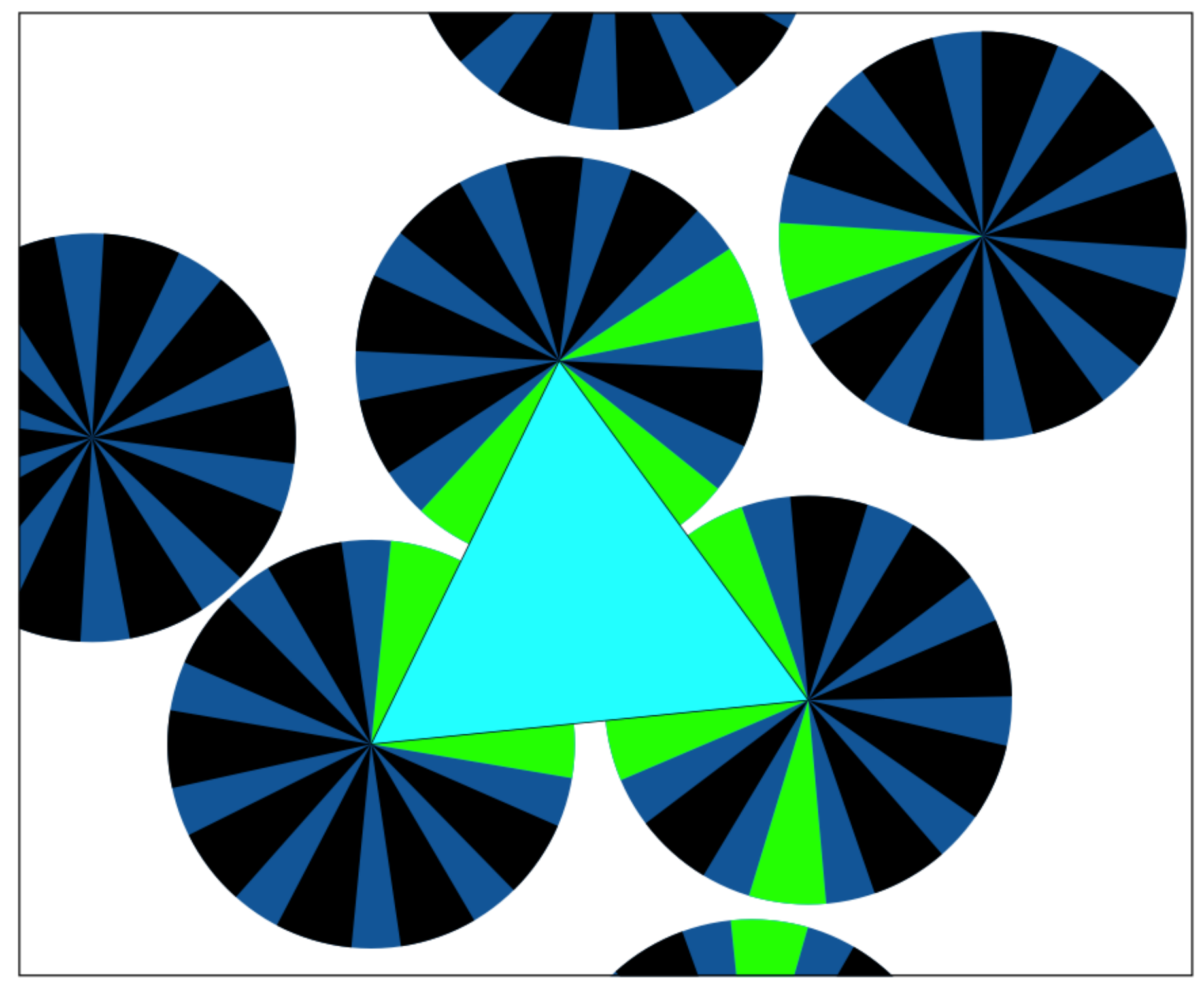} 
   \caption{\label{fig_model} Figure illustrating the nature of the disc model. Sticky patches (black) of opening angle $2 w$ are shown green when engaged in pairwise interactions, which come with energetic reward $-\epsilon\, \kt$. In figures we label polygons corresponding to the shortest possible convex closed loops that can be drawn by joining energetically-interacting disc centers. Each of the three sides of this triangle is a line joining the centers of pairwise-interacting discs.}
\end{figure}

In many figures we show polygons corresponding to the shortest possible convex closed loops that can be drawn by joining energetically-interacting disc centers (see \f{fig_model}), according to the following color scheme: polygons with 3 sides are cyan, polygons with 4 sides are red, polygons with 5 sides are magenta, polygons with 6 sides are dark blue, and polygons with 7 sides are light green (in some figures, only polygons of size 3 and 4 are shown).

\section{Cooling simulations}
\label{sec_cooling}

We did `cooling' simulations within square boxes of edge length 40$a$, where $a$ is the disc diameter. The box was initially populated with randomly dispersed and oriented non-overlapping discs, at 25\% packing fraction. We set the initial disc bond energy parameter $\epsilon$ to 2. Every $5 \times 10^8$ simulation steps (attempted individual moves, not `sweeps') we `cooled' the system by increasing $\epsilon$ by 0.1 (as in the simulations done to produce \f{fig3}(a) and \f{fig_loop_counts}) or by 0.03 (as in the simulations done to produce \f{fig2}(c), \f{fig_array}, and \f{fig_array2}). 

At each step of the simulation we attempted a grand-canonical move with probability $(1+N)^{-1}$, where $N$ is the instantaneous number of discs on the substrate. We attempted an on-substrate move otherwise. If a grand-canonical move was attempted then we proposed with equal likelihood a particle insertion or deletion, and accepted these proposals with probabilities
\beq
\label{rate1}
p_{\rm insert}(N \to N+1) =\min \left(1, \frac{V}{N+2}  {\rm e}^{ \beta \mu - \beta \Delta E } \right)
\eeq
or 
\beq
\label{rate2}
p_{\rm delete}(N \to N-1) = \min \left(1, \frac{N+1}{V} {\rm e}^{ -\beta \mu - \beta \Delta E }\right).
\eeq
Here $\mu$ is a chemical potential term, $V$ is the (dimensionless) box area, and $\Delta E$ is the energy change (due to disc-disc interactions) resulting from the proposed move. The combinations $N+2$ and $N+1$ that appear the right-hand sides of \eqq{rate1} and \eqq{rate2} become $N+1$ and $N$ if grand-canonical moves are proposed with fixed probability. The chemical potential $\mu$ was chosen so that the disc packing fraction in the absence of attractive interactions was 25\%. 

For on-substrate moves we used either standard Metropolis single-particle dynamics or the virtual-move Monte Carlo procedure described in the appendix of Ref.\csupp{whitelam2009rcm}, or both. We found that the qualitative nature of the first condensed phase to self-assemble from `solution' (i.e. the low-occupancy gas of discs that was the starting point for each cooling simulation) did not depend on which combination of these algorithms we used. \f{fig2}(c), \f{fig_array}, and \f{fig_array2} were obtained using the Metropolis algorithm only, and \f{fig3}(a) was obtained using a 50:50 mixture of Metropolis and virtual moves. Cooling simulations were run for of order $10^{11}$ steps.
\begin{figure}[ht] 
   \centering
 \includegraphics[width=\linewidth]{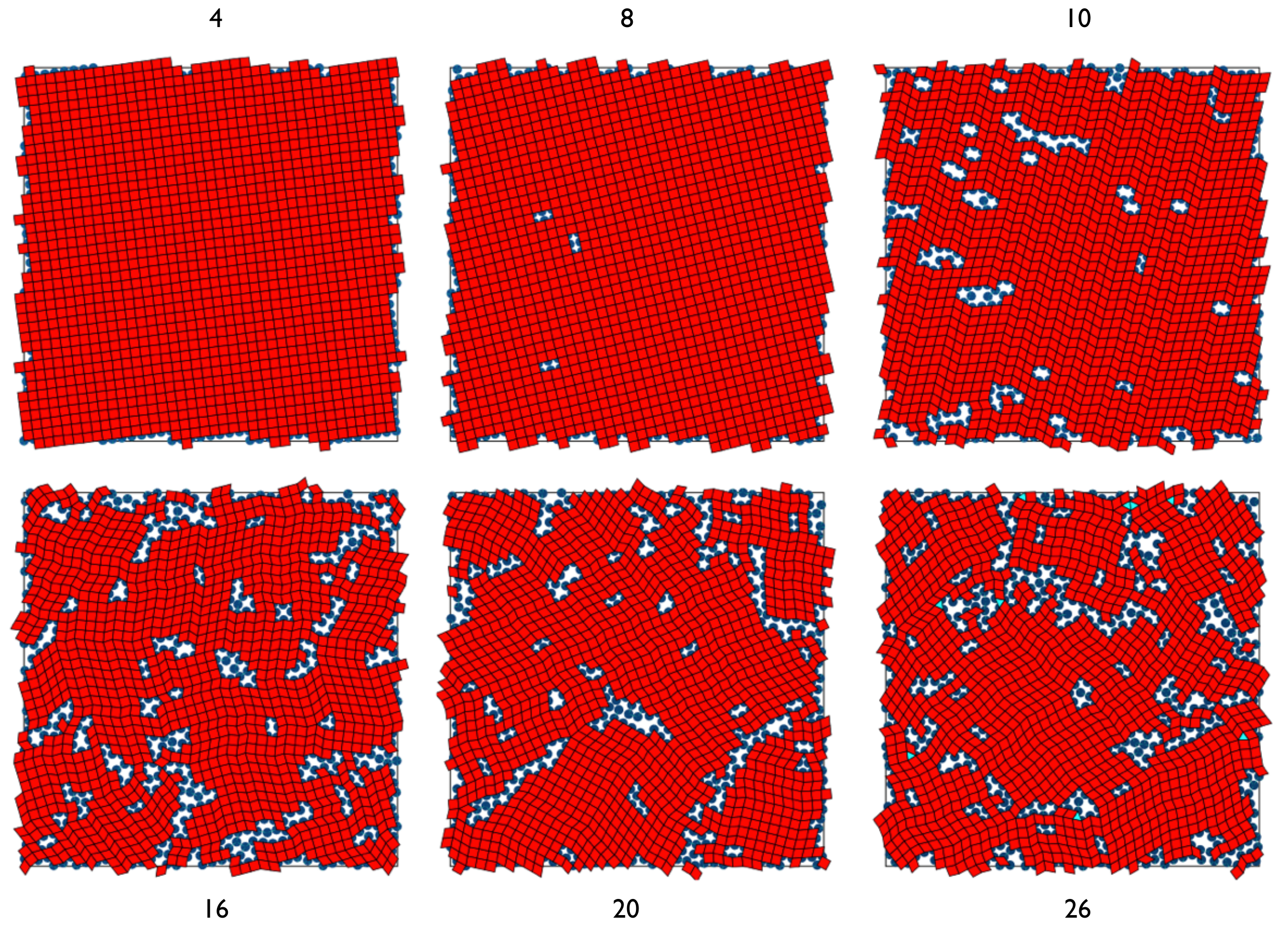} 
   \caption{\label{fig_array} Simulation boxes showing tilings obtained upon cooling collections of $M$-patch discs of patch half-angle $w=2^\circ$, where $M$ is labeled on the panels. All values of $M$ are those for which the simple geometric argument of the main text predicts that unstrained rhombus tilings can form; these simulations show that such tilings indeed self-assemble spontaneously. Polygons up to size 4 (red) are shown.}
\end{figure}

\begin{figure}[ht] 
   \centering
 \includegraphics[width=\linewidth]{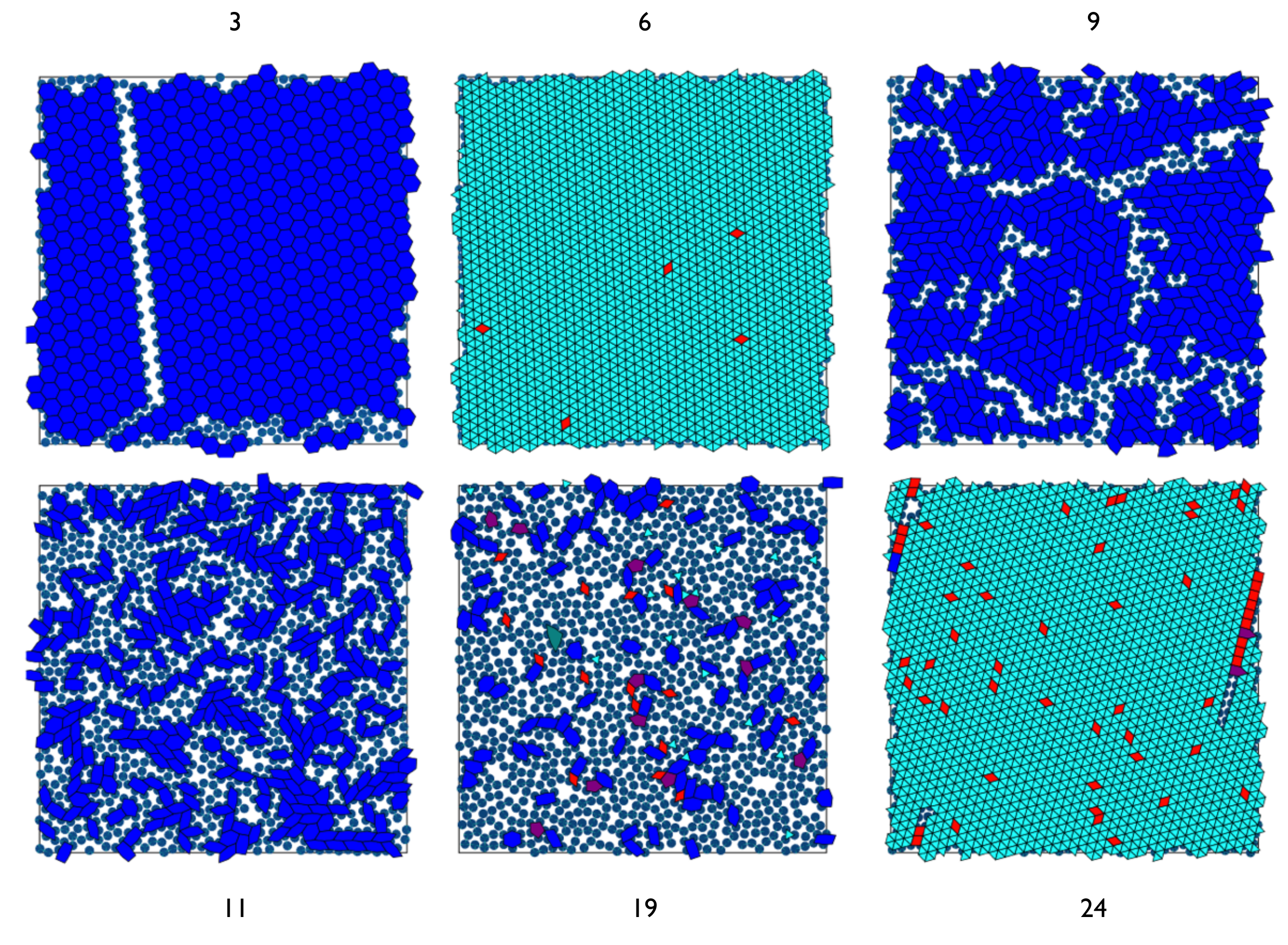} 
   \caption{\label{fig_array2} Simulation boxes showing tilings obtained upon cooling collections of $M$-patch discs of patch half-angle $w=2^\circ$, where $M$ is labeled on the panels. All values of $M$ are those for which the simple geometric argument of the main text predicts that unstrained rhombus tilings will not form, either because $M$ is not even, or because it is a multiple of 6 (for which the triangular tiling is energetically preferred). Simulations confirm this expectation: no rhombus tilings have self-assembled in these examples. Note that 9-patch discs self-assemble into a tiling of irregular hexagons. Polygons of size 3 (cyan), 4 (red), 5 (magenta), 6 (dark blue), and 7 (light green) are shown.}
\end{figure}

\begin{figure}[ht] 
   \centering
 \includegraphics[width=\linewidth]{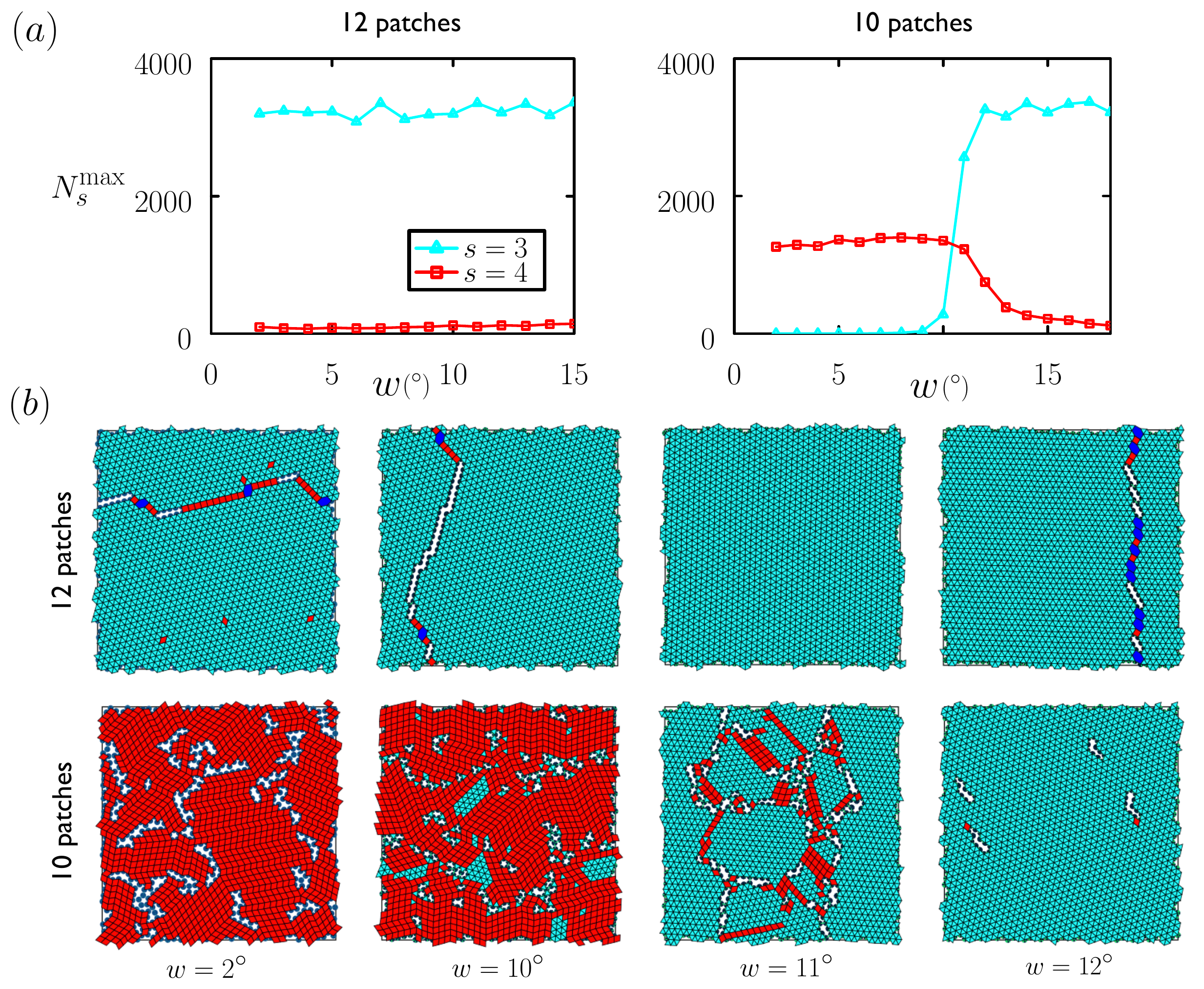} 
   \caption{\label{fig_loop_counts} Emergent rhombus tilings display polymorphism (see \f{fig3}). (a) Left: Cooling collections of 12-patch discs results in self-assembly of the triangle phase, regardless of the orientational specificity of interactions: strain-free triangular tilings are possible in the limit of infinitesimal patch width -- in part because 12 is a multiple of 6 --  and triangular tilings are also expected in the isotropic limit ($w=\pi/12=15^\circ$). Simulations are consistent with these expectations: we plot the largest number $N_s^{\rm max}$ of polygons (drawn atop networks) having $s$ sides that were seen over the course of long cooling simulations; appreciable numbers of triangles only were seen (the wobbles in the blue line result from the different numbers of grain boundaries seen simulation-to-simulation, owing to the stochastic nature of nucleation and growth). Right: As plotted in \f{fig3}(a), cooling 10-patch discs result in 4-gons (rhombuses) when interactions are orientationally specific -- in part because 10 is a multiple of 2 and not of 6 --  and of the triangle phase as interactions approach the isotropic limit. Between these limits the two phases can coexist and interconvert, with rhombuses favored by entropy and triangles by energy. Panel (b) shows snapshots taken from cooling simulations of 10- and 12-patch discs, for different values of $w$. For 12-patch discs we observe only triangular tilings. For 10-patch discs, for certain values of $w$, rhombuses and triangles abut.}
\end{figure}

\section{Rhombus tile self-assembly}
\label{sec_rhombus}

The right-hand snapshot of \f{fig2}(a) was produced using the off-lattice rhombus model of Ref.\csupp{whitelam2012random}, and the procedure used to produce Fig. 2 of that reference. In brief, a compact cluster of rhombus tiles possessing internal angles $2\pi/5$ and $3 \pi/5$ was grown in the grand-canonical ensemble using umbrella sampling. The rhombus-rhombus interactions and chemical potential of the system were chosen so that a low-density gas of tiles was metastable with respect to a condensed tiling. An harmonic constraint $\propto (N-N_0)^2$ was applied to the size $N$ of the largest cluster in the system, and the `center' $N_0$ of that constraint was slowly increased in order to promote the nucleation and growth of a cluster of about 1000 tiles. The environment shown in \f{fig2}(a) is taken from within that cluster.

\section{Rhombus-triangle stability analysis for 10-patch discs}
\label{sec_stability}

{\em Analytic theory}. To make \f{fig3}(b) we calculated the bulk free-energy difference between rhombus and triangle phases of 10-patch discs within a mean-field approximation\csupp{geng2009theory}. In this approximation one assumes that a bulk phase can be described by focusing on a single particle, and treating the environment of that particle in an implicit way. Let us imagine that a particle's accessible configuration space -- the set of angles and positions it can adopt -- is discretized, consisting of $K$ microstates labeled $\alpha = 1,2,\dots,K$. In each microstate, let the sum of the particle's interactions with its environment be $E_{\alpha}$. The probability of observing microstate $\alpha$ in thermal equilibrium is then $p_\alpha = Z^{-1} {\rm e}^{-\beta E_\alpha}$, where $Z= \sum_\alpha  {\rm e}^{-\beta E_\alpha}$. We assume zero pressure, an assumption intended to model simulations in which domains of triangles and rhombuses have free boundaries.

The bulk free energy of a (noninteracting) collection of such particles is $f = U -TS$, with $U = \frac{1}{2}\langle E_\alpha \rangle$ (the factor of $1/2$ appears because bond energies are shared between two particles) and $S =- k_{\rm B} \langle \ln p_\alpha \rangle$. Note that thermal averages are defined by $\langle \cdot \rangle = Z^{-1} \sum_\alpha \left( \cdot \right) {\rm e}^{-\beta E_\alpha}$. We therefore have 
\bea
\label{eq_mf}
f &=& \frac{1}{2}\langle E_\alpha \rangle + \kt  \langle \ln p_\alpha \rangle \nonumber \\
&=& \frac{1}{2}\langle E_\alpha \rangle + \kt  \langle \ln Z^{-1} {\rm e}^{-\beta E_\alpha} \rangle \nonumber \\
 &=&  -\frac{1}{2}\langle E_\alpha \rangle -\kt \ln Z.
\eea
So far this analysis is generic. To specialize it to the $M$-patch disc model considered in this paper, we note that a particle's energetic environments $\alpha$ are defined only by the number of engaged bonds $n = 0,1,2,\dots\,6$ that the particle possesses (steric constraints set the largest possible number of engaged bonds to be 6). Given that the bond interaction energy is $-\epsilon \, \kt$, we can write \eqq{eq_mf} as
\beq
\label{eqf1}
\beta f_{\triangle} = \frac{1}{2} \frac{\sum_{n=0}^6 \epsilon n \, g_{\triangle}(n) {\rm e}^{\epsilon n}}{\sum_{n=0}^6 g_{\triangle}(n) {\rm e}^{\epsilon n}}-\ln \sum_{n=0}^6 g_{\triangle}(n) {\rm e}^{\epsilon n},
\eeq
for the triangle phase of discs, and as
\beq
\label{eqf2}
\beta f_{\lozenge} = \frac{1}{2} \frac{\sum_{n=0}^6 \epsilon n \, g_{\lozenge}(n) {\rm e}^{\epsilon n}}{\sum_{n=0}^6 g_{\lozenge}(n) {\rm e}^{\epsilon n}}-\ln \sum_{n=0}^6 g_{\lozenge}(n) {\rm e}^{\epsilon n},
\eeq
for the rhombus phase. In these expressions the symbol $g_\triangle(n)$ stands for the total configuration space volume within which a particle in a triangle phase possesses $n$ engaged bonds; $g_\lozenge(n)$ is the corresponding symbol for the rhombus phase. The terms $g$ therefore quantify the rotational and vibrational entropy possessed by particles in particular environments. 

Our aim is to compute the bulk free-energy difference between phases, i.e.
\beq
\label{complex}
\beta \Delta f = \beta f_{\triangle}-\beta f_{\lozenge}.
\eeq
We found that this difference is dominated, for the range of parameters we used in our simulations, by the 6-coordinated environments of the triangle phase and the 4-coordinated environments of the rhombus phase, meaning that $\sum_n g_\triangle(n) {\rm e}^{\epsilon n} \approx g_\triangle(6) {\rm e}^{6 \epsilon}$ and $\sum_n g_\lozenge(n) {\rm e}^{\epsilon n} \approx g_\lozenge(4) {\rm e}^{4 \epsilon}$. To this approximation, Equations \eq{eqf1} and \eq{eqf2} become
\beq
\beta f_{\triangle} \approx -3 \epsilon - \ln g_{\triangle}(6)
\eeq
and
\beq
\beta f_{\lozenge} \approx -2 \epsilon - \ln g_{\lozenge}(4),
\eeq
and \eqq{complex} becomes
\beq
\label{simple}
\beta \Delta f = -\epsilon + \ln \left(\frac{g_{\lozenge}(4)}{g_{\triangle}(6)} \right),
\eeq
which separates in an obvious way into energetic and entropic terms. 
\begin{figure}[ht] 
   \centering
 \includegraphics[width=0.9\linewidth]{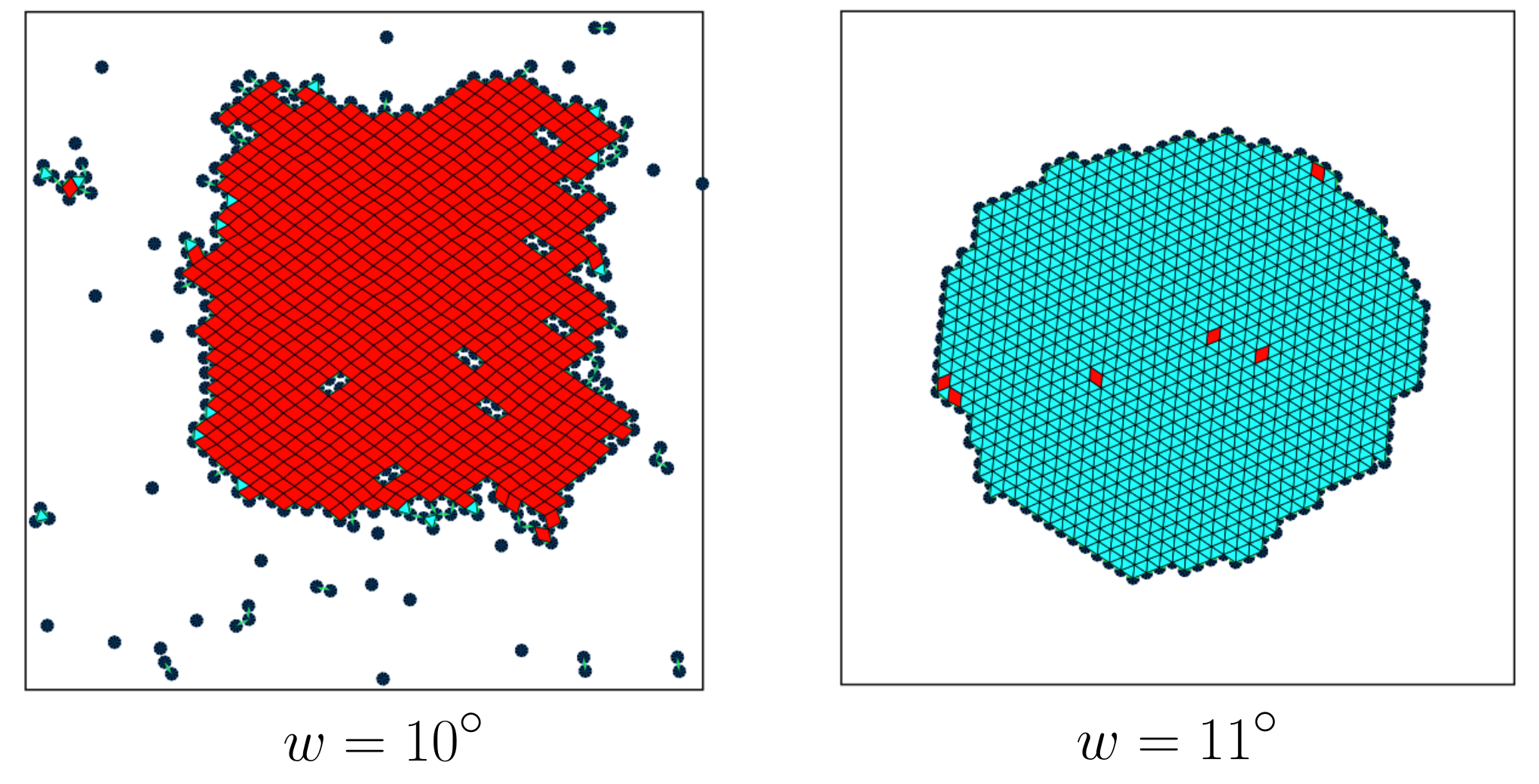} 
   \caption{\label{fig_phase_space_start} Typical initial conditions for the configuration space calculations described in \s{sec_stability}, showing rhombus-phase and triangle-phase clusters with free boundaries.}
\end{figure}

{\em Numerical computation of vibrational and rotational entropy.} We chose to calculate the vibrational and rotational entropy of discs in condensed phases numerically, for convenience, because the lattice structure of the triangle phase changes continuously with $w$ (we note that Ref.\csupp{mao2013entropy} describes an analytic method for calculating the vibrational and rotational entropy of periodic phases of patchy particles). This entropy results from the number of configurations $g_\lozenge(n)$ and $g_\triangle(n)$ possessed by particles in particular environments. We considered 10-patch discs, and worked in the $NVT$ ensemble in order to allow the formation of clusters of the rhombus and triangle phases with free boundaries (see \f{fig_phase_space_start}), so avoiding a possible bias in favor of either phase resulting from the shape of the simulation box. We therefore worked effectively at zero pressure. We simulated 1000 discs within a square simulation box at a packing fraction of 30$\%$. We ran three independent simulations for each of various fixed values of the patch width parameter $w$; these values ranged from $9.8^\circ$ to $12.0^\circ$, in increments of $0.1^\circ$. To generate the rhombus phase, we placed particles `by hand' into rhombus formation, forming a cluster of size $30\times 30$ (the remaining 100 particles were dispersed at random, subject to there being no disc-disc overlaps, throughout the simulation box). We relaxed these structures using short simulations of $10^5$ Monte Carlo sweeps (moves per disc) in which we chose to apply either the Metropolis algorithm (with likelihood 90\%), the virtual-move Monte Carlo algorithm (5\%), or the nonlocal algorithm described in the appendix of Ref.\csupp{whitelam2010control} (5\%). Nonlocal algorithms are a convenient way of relaxing structures made from strong bonds and overcoming the slowness of diffusion; see e.g. Ref.\csupp{chen2001improving}. We set $\epsilon=5$. To generate initial configurations of the triangular phase -- actually phases: they differ in structure as $w$ changes -- we took the right-hand configuration shown in \f{fig3}(c), which resulted from a simulation carried out in the $NVT$ ensemble at 30\% packing fraction and a bond strength $\epsilon=7$. We carried out a set of short Monte Carlo simulations starting from this configuration, decreasing $w$ by small amounts and relaxing the new triangle phase that resulted from each change, until we had generated a triangle phase for patch width $w=10.3^\circ$. This latter configuration was the starting point for a set of three independent simulations done at each of the values of $w$ in the range $w=9.8^\circ$ to $12.0^\circ$, in increments of $0.1^\circ$. We used a similar relaxation procedure as for the rhombus phase, although of greater length ($2 \times 10^6$ sweeps), so as to give each distinct triangular phase time to emerge. We also simulated the triangle phases using using a bond strength of $\epsilon=7$, rather than the $\epsilon=5$ used for the rhombuses.

From each simulation we took the final-time configuration, and used these static configurations to calculate the degeneracy terms $g(n)$ required as input to the mean-field theory. For a given simulation, this calculation proceeded as follows. We considered only `bulk' rhombus-phase or triangle-phase particles, which we defined as $k$-fold coordinated particles possessing {\em only} $k$-fold coordinated neighbors, $k$ being 4 for the rhombus phase and 6 for the triangle one. For each such particle we considered its configuration space to be a grid of $(501)^3$ distinct Cartesian coordinates and orientation angles, of total extent $a \times a \times 2 \pi/10$, centered on the position and orientation the disc possessed in the static configuration. We calculated the number ${\cal N}(n)$ of these grid points at which the disc had $n$ engaged patches and no overlaps; see \f{fig_cell} for examples of such configurations. We replaced the disc in its original position, and repeated this calculation for all bulk discs. We averaged the quantities ${\cal N}(n)$ over all bulk discs (in all three independent simulations done at each value of $w$), thereby averaging over of order $10^3$ distinct microscopic environments.

We then note that ${\cal N}(n) \propto g(n)$, with the constant of proportionality being unimportant for the purposes of computing the difference of Equations~(\ref{eqf1}) and~(\ref{eqf2}) (because that constant contributes the same additive constant to each equation). 
\begin{figure}[ht] 
   \centering
 \includegraphics[width=0.9\linewidth]{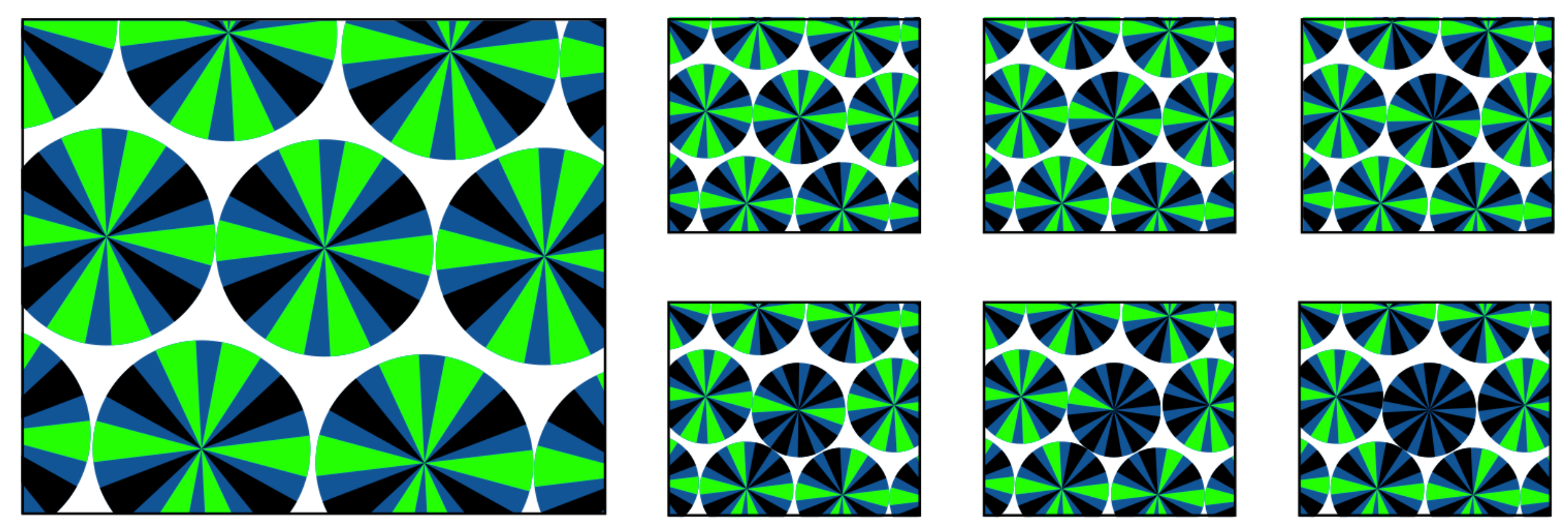} 
   \caption{\label{fig_cell} A visual representation of the numerical procedure used to compute the rotational and vibrational entropy of bulk phases (related to the factors $g(n)$). Here we show snapshots of the triangle phase, for $w=11^\circ$. The seven pictures show one configuration in which the central disc has $n$ engaged patches, where $n$ runs from 0 to 6. The total number of configurations of each type is proportional to $g(n)$.}
\end{figure}

The result of these calculations, fed into \eqs{complex} and \eq{simple}, can be seen in \f{fig3}(b). Here we plot the value $\epsilon_0$ of the bond strength, as a function of patch width $w$, at which bulk rhombus and triangle phases are equal in free energy (i.e. we set the left-hand sides of \eqs{complex} and \eq{simple} to zero, and we calculated the resulting value of $\epsilon$, which we call $\epsilon_0$). In \f{fig3}(b), the line labeled `full' was obtained from \eqq{complex}, and the line labeled `basic' was obtained from \eqq{simple}. Note that this plot predicts that the rhombus phase, which was simulated at $\epsilon=5$, is less stable than the triangle phase, at that value of $\epsilon$, when $w\gtrsim 11^\circ$. It also predicts that the triangle phase, which was simulated at $\epsilon=7$, is less stable than the rhombus phase, at that value of $\epsilon$, when $w \lesssim 10^\circ$. Our simulations are consistent with this prediction: in those regimes we saw rhombus and triangle phases begin to convert to the other phase. Such instability highlights one convenient feature of this method: the phase in question need not be stable, or even particularly long-lived, in order to have its vibrational properties calculated (one needs only a static snapshot of the material). A second convenient feature of this method is that it furnishes the free-energy difference between phases {\em away} from coexistence, as well as allowing one to estimate the points in configuration space at which phases are equal in free energy.

Note that the value of $\epsilon$ at which these two condensed phases are unstable with respect to dissolution depends, given $w$, on the value of the disc packing fraction. For the 30\% used for these simulations, and for $w=11^\circ$, the rhombus phase begins to dissolve into a liquidlike phase for $\epsilon$ less than about 4.75. We also found that $g_\triangle(6) \propto (w-w_0)^2$, where $w_0 \approx 9.7^\circ$, indicating that the triangle phase cannot exist for $w \lesssim 9.7^\circ$.

{\em Sources of error.} The framework we have used to calculate the free-energy difference between rhombus and triangle phases is a mean-field one, and so one systematic error we incur results from the neglect of particle correlations that was assumed in order to produce \eqq{eq_mf}. Within mean-field theory, we cannot estimate how large this error might be. Another potential source of systematic error in this calculation results from the fact that we simulated the rhombus phase at bond strength $\epsilon=5$ and the triangle phase at $\epsilon=7$ for convenience: for some values of $w$, simulating these phases using the same value of $\epsilon$ makes one phase convert to the other so rapidly that it is hard even to `locally relax' it before it disappears. We then assumed that the values of $g(n)$ so obtained are independent of $\epsilon$, allowing us to vary $\epsilon$ freely in Equations~(\ref{eqf1}) and~(\ref{eqf2}). This assumption is based on the fact that both values of $\epsilon$ are large enough to ensure that simulations of rhombus or triangle phases produce large expanses of fully-connected network, and that the properties of connected networks, for the square-well potential we have used, depend little on the precise value of $\epsilon$ used to establish that connectivity. Although we do expect transient broken bonds within networks to influence the microscopic environments of surrounding particles, we expect this influence to be relatively minor for $\epsilon$ several times larger than $\kt$. The agreement between the expressions \eqq{simple} and \eqq{complex} seen in \f{fig3}(b) suggests that this assumption is a reasonable one.

We have estimated the statistical error in our calculation in the following way. The quantities $g_\triangle(n)$ and $g_\lozenge(n)$ result from an average over of order $10^3$ microscopic environments. We calculated the standard deviations $\sigma_\triangle(n)$ and $\sigma_\lozenge(n)$ of the distributions from which these averages were calculated, and estimated the error bars shown in \f{fig3}(b) by evaluating \eqq{simple} with the replacements $g_\triangle(6) \to g_\triangle(6) \pm \sigma_\triangle(6)/2$ and $g_\lozenge(4) \to g_\lozenge(4) \mp \sigma_\lozenge(4)/2$ (the upper and lower signs giving us the top and bottom of each error bar). For all values of $w$ considered, the error in the phase boundary $\epsilon_0(w)$ is roughly $\kt$. 
 
Despite the many possible sources of error in this calculation, the mean-field estimate derived from \eqs{complex} and \eq{simple}, for $w=11^\circ$, agrees with the results of direct coexistence simulations with reasonable precision. The small symbols shown in a vertical line at $w=11^\circ$ on \f{fig3}(b) describe the results of direct coexistence simulations done in the $NVT$ ensemble using Metropolis (90\%), virtual-move (5\%) and nonlocal (5\%) algorithms. (The Metropolis algorithm facilitates single-particle rotations that enable conversion between triangle and rhombus phases. Within the virtual-move algorithm, the rate of rotating a single particle in such a way that it breaks bonds with neighboring particles diminishes with each bond broken; by contrast, within the Metropolis algorithm, rotational moves that break many bonds can happen without penalty if they result in the creation of at least that many bonds.) These simulations were begun from the middle configuration shown in \f{fig3}(c). A square (resp. triangle) symbol denotes eventual transformation to the rhombus (resp. triangle) phase, from which we estimate the phase boundary $\epsilon_0$ to be between 5 and 5.1. The phase boundary derived from \eqq{complex} is $ \epsilon_0(11^\circ) = (4.8 \pm 0.6)\, \kt$, consistent with the direct-coexistence result.

\section{Archimedean tilings}

\begin{figure}[b!] 
   \centering
 \includegraphics[width=0.5\linewidth]{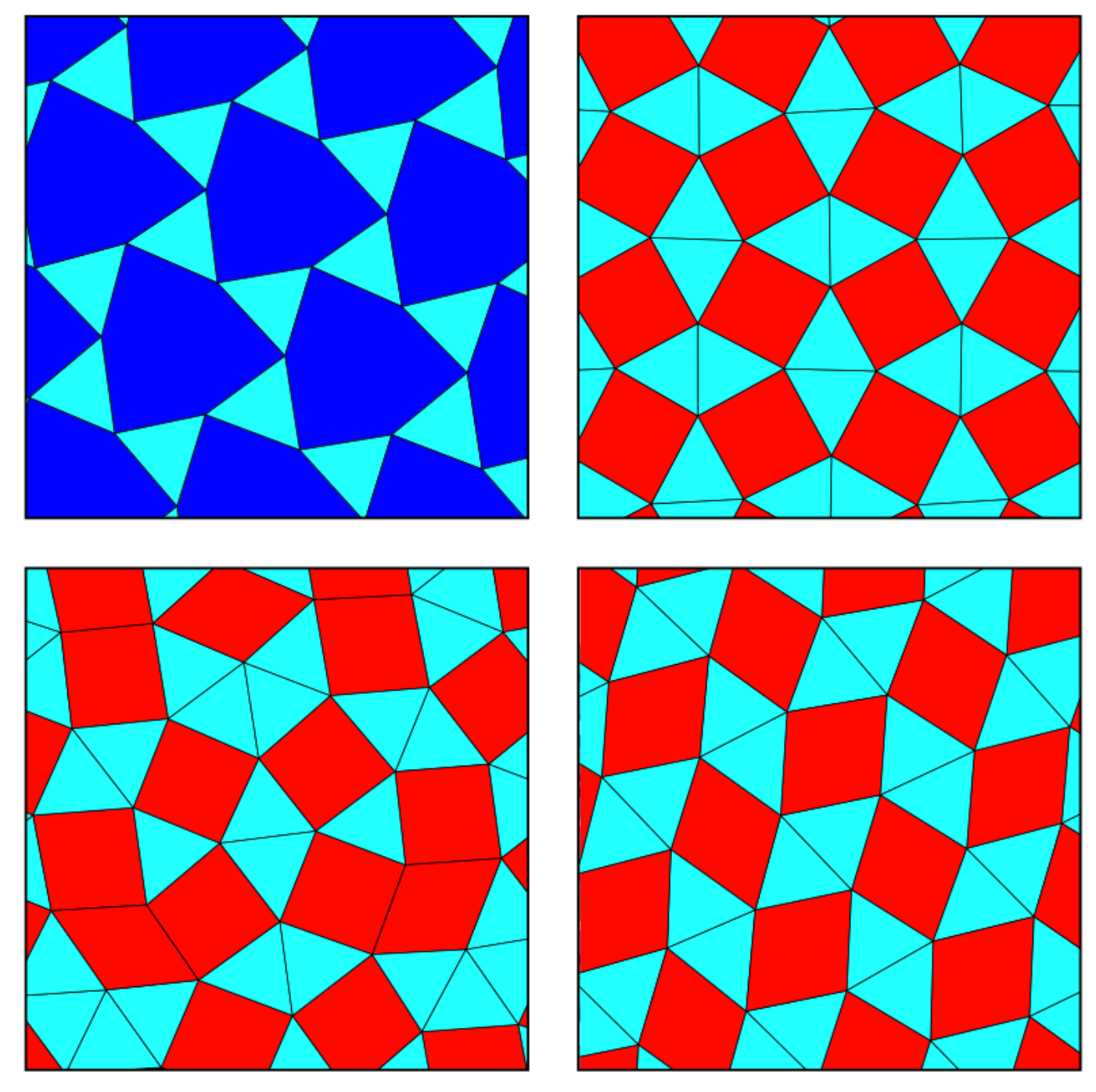} 
   \caption{\label{fig5} Networks equivalent to Archimedean tilings and distortions thereof self-assemble from patchy discs with $M$ odd and $w$ sufficiently wide (see\csupp{doye2007controlling,van2012formation,PhysRevLett.110.255503}). Clockwise from top left: `fish scale' irregular trihexagonal tiling ($M=5,w=10^\circ$); snub square tiling ($M=5,w=13^\circ$); rhombus analog of the snub square tiling ($M=11,w=8.1^\circ$); and nonperiodic square-triangle tiling ($M=5,w=25^\circ$; see e.g. Ref.\csupp{doye2007controlling}).}
\end{figure}

As discussed in the main text, self-assembly of Archimedean tilings can be achieved using irregular patch placement\csupp{antlanger2011stability}, or an odd number of suitably wide patches\csupp{doye2007controlling,van2012formation,PhysRevLett.110.255503} (or the equivalent effective coordination\csupp{ecija2013five,millan2014self}). In \f{fig5} we show networks self-assembled from patchy discs with odd-numbered rotational symmetry. Self-assembly of the snub square tiling and the nonperiodic square-triangle tiling has been demonstrated previously (e.g. \csupp{doye2007controlling,doppelbauer2010self,ecija2013five}); assembly of the other two patterns (irregular variants of Archimedean tilings), so far as we know, has not.

%\bibliography{bib}

\end{document}